\documentclass[aps,
preprint,
showpacs]{revtex4-1}
\usepackage{amssymb}
\usepackage{amsfonts}
\usepackage{amsmath}
\usepackage{mathrsfs}
\usepackage{graphicx}
\usepackage{flushend}

\begin{document}

\title{A New Improved Energy-Momentum Tensor and Its Possible Role in Gravity}

\author{Tao Lei}
\author{Zi-Wei Chen}
\author{Zhen-Lai Wang}
\author{Xiang-Song Chen}
\email{cxs@hust.edu.cn}

\affiliation{School of Physics and
MOE Key Laboratory of Fundamental Quantities Measurement,
Huazhong University of Science and Technology, Wuhan 430074, China}

\date{\today}

\begin{abstract}
Motivated by a special consideration in quantum measurement, we present a new improved energy-momentum tensor. The new tensor differs from the traditional canonical and symmetric ones, and can be derived as N\"other current from a Lagrangian with second derivative. We also attempt to construct a gravitational coupling in such a way that the new energy-momentum tensor becomes the source of the gravitational field. The theory we obtain is of an Einstein-Cartan type, but derived from a minimal coupling of a Lagrangian with second-derivative, and leads to additional interaction between torsion and matter, including the scalar field. For the scalar field, the theory can also be derived in the Riemann space-time by a non-minimal coupling. Our study gives hint on more general tests of general relativistic effects.
\end{abstract}
\maketitle

\section{Introduction}
Being the conserved current associated with the symmetry of space-time translation, energy-momentum tensor is among the most fundamental quantity in both classical and
quantum physics \cite{Lead14}, and is particularly important for the formulation of a consistent gravitational theory \cite{Hehl76, Coleman70,Hehl75}. There are two popular expressions of energy-momentum tensor. One is the canonical energy-momentum tensor, derived from N\"other's theorem:
\begin{eqnarray}\label{Tcano}
{T}_{cano}^{\mu\nu}=-\frac{\partial\mathscr{L}_{\rm st}(\varphi,\partial\varphi)}{\partial(\partial_\mu\varphi)}{\partial}^\nu\varphi+\eta^{\mu\nu}\mathscr{L}_{\rm st},
\end{eqnarray}
where $\mathscr{L}_{\rm st}(\varphi, \partial \varphi)$ is the standard expression of matter Lagrangian in terms of the
field $\varphi$ and its first derivative, and the Minkowski metric tensor $\eta_{\mu\nu}$ has signature $($$-$$+$$+$$+$$)$. The other is the symmetric energy-momentum tensor, known as the Belinfante tensor:
\begin{eqnarray}
{T}_{symm}^{\mu\nu}={T}_{cano}^{\mu\nu}-\frac{i}{2}\partial_\rho[\frac{\partial\mathscr{L}_{\rm st}}{\partial(\partial_\rho\varphi_a)}(\Sigma^{\mu\nu})_{ab}\varphi^b
+\frac{\partial\mathscr{L}_{\rm st}}{\partial(\partial_\nu\varphi_a)}(\Sigma^{\mu\rho})_{ab}\varphi^b
+\frac{\partial\mathscr{L}_{\rm st}}{\partial(\partial_\mu\varphi_a)}(\Sigma^{\nu\rho})_{ab}\varphi^b],
\end{eqnarray}
where $\Sigma^{\mu\nu}$ are a set of matrices satisfying the algebra of the homogeneous Lorentz group:$[\Sigma^{\mu\nu},\Sigma^{\rho\sigma}]=2i(\eta^{\mu[\sigma}\Sigma^{\rho]\nu}-\eta^{\nu[\rho|}\Sigma^{\mu|\sigma]})$. In this paper, $[~]$/$(~)$ means antisymmetrization/symmetrization, and indices inside $|~|$ are excluded from symmetrization or antisymmetrization. Applied to gravitational theory, the symmetric energy-momentum tensor leads to Eenstein's general relativiy,
and the canonical energy-momentum tensor leads to Eenstein-Cartan theory \cite{Hehl75, Hehl89}.

In this paper, we present a new type of energy-momentum tensor, and apply it to explore what theory we get by setting the new  energy-momentum tensor as the gravitational source. To show the difference, let us first
give the explicit expression of our new energy-momentum tensor. It is
\begin{equation}\label{Tnew}
T^{\mu\nu}_{\rm new}=-\frac{\partial\mathscr{L}_{\rm st}}{\partial(\partial_\mu\varphi)}\overleftrightarrow{\partial}^\nu\varphi
=-\frac{1}{2} \big( \frac{\partial\mathscr{L}_{\rm st}}{\partial(\partial_\mu\varphi)}{\partial}^\nu\varphi
-\partial^\nu\frac{\partial\mathscr{L}_{\rm st}}{\partial(\partial_\mu\varphi)}\varphi \big),
\end{equation}
where $\overleftrightarrow{\partial}^\nu=\frac{1}{2}(\overrightarrow{\partial}^\nu-\overleftarrow{\partial}^\nu)$. The explicit forms of our new energy-momentum tensors for the scalar, Dirac, and vector fields are:
 \begin{subequations}
\begin{align}
{T}_{\rm new}^{\mu\nu}&=\partial^\mu\phi\overleftrightarrow{\partial}^\nu\phi
=\frac{1}{2}(\partial^\mu\phi\partial^\nu\phi-\phi\partial^\mu\partial^\nu\phi),\\
{T}_{\rm new}^{\mu\nu}&=-i\overline{\psi}\gamma^\mu\overleftrightarrow{\partial}^\nu\psi
=\frac{i}{2}[-\overline{\psi}\gamma^\mu\partial^\nu\psi+(\partial^\nu\overline{\psi})\gamma^\mu\psi],\\
{T}_{\rm new}^{\mu\nu}&=F^{\mu\rho}\overleftrightarrow{\partial}^{\nu}A_\rho
=\frac{1}{2}(F^{\mu\rho}\partial^{\nu}A_\rho-\partial^{\nu}F^{\mu\rho}A_\rho).%
\end{align}
\end{subequations}

\section{Derivation of the new expression}
Our motivation to re-examine the expression of energy-momentum tensor is that
the conserved current is not uniquely determined by the conservation law, which can only prescribe total
conserved charge, and much debate arose\cite{Obuk15,Tint11,Tint13}.
On the other hand, the conserved current {\em densities} do have independent physical meanings. The most familiar example is that the energy-momentum tensor acts as the source of gravitational field in gravitational theory. In this paper, we seek to set a constraint on the energy-momentum.

Our consideration is that \textit{if a quantum wave is in mutual eigen-state of more than one physical observables,
and a simultaneous measurement of these observables can be performed, then the currents associated with these
observables must be proportional to each other.}

The hint on such correlation of currents comes from classical particles: When one catches a classical particle, one
catches all its physical quantities: charge, energy, momentum, etc. Thus, for a beam of identical particles with the
same energy $\varepsilon$ and momentum $p_j$ for each particle, the  energy flux density $\vec {\cal K}_0$ must be
proportional to the momentum flux density $\vec {\cal K}_j$:
\begin{equation}\label{classical}
\dfrac {\vec {\cal K}_0}{\varepsilon}= \dfrac {\vec {\cal K}_j}{p_j} =\vec {\cal K}_n,
\end{equation}
with $\vec {\cal K}_n$ the flux density of particle number.
The case will be trivial if all components of $p_j$ are identical for the particles, but non-trivial cases
can be designed if just one or two components of $p_j$ are set identical. The same remark applies to the discussion
of quantum wave below.

By the assumption of quantum measurement, when a quantum wave collapses to a local spot, all its physical
quantities will localize simultaneously to that same spot. In this way, a quantum wave should exhibit similar correlation
of currents as for classical particles: If the wave is in mutual eigen-state of energy $\varepsilon$ and momentum
$p_j$, then the density of energy flow $T^{i0}$ and the density of momentum flow $ T^{ij}$ must satisfy
a constraint similar to Eq. (\ref{classical}):
\begin{equation}\label{quantum}
\dfrac {T^{i0}}{\varepsilon}= \dfrac{T^{ij}}{p_j},
\end{equation}
so that one can have
\begin{equation}\label{measure}
\dfrac {T^{i0}\cdot dS_i}{\varepsilon}= \dfrac{T^{ij}\cdot dS_i}{p_j}=\dfrac{dN}{dt},
\end{equation}
where $N$ is the number of particles received at the surface element $d\vec S$.

It is interesting and surprising that the conventional expressions of energy-momentum tensor do not show the
above correlation.
For example, taking the canonical expression in Eq. (\ref{Tcano}), and making use of the eigen-state assumption
$\partial^0\varphi=i \varepsilon \varphi$,
$\partial^j\varphi=i p_j \varphi$, we have:
\begin{subequations}
\begin{align}
T_{\rm cano}^{i0}&\rightarrow- i \varepsilon\dfrac{\partial{\mathscr  L}_{\rm st}}{\partial(\partial_i\varphi)}\varphi,
\label{i0} \\
T_{\rm cano}^{ij}&\rightarrow- i p_j\dfrac{\partial{\mathscr  L}_{\rm st}}{\partial(\partial_i\varphi)}\varphi+
\delta_{ij}{\mathscr  L}_{\rm st}.\label{ij}
\end{align}
\end{subequations}
This satisfies the constraint in Eq. (\ref{quantum}) for the transverse momentum flow, namely $T^{ij}$ with $i\neq j$.
But for the longitudinal momentum flow $T^{jj}$,  the Lagrangian term in Eq. (\ref{ij}) makes a trouble for Eq.
(\ref{quantum}), except for the Dirac field $\psi$ which has
$ {\mathscr  L}^\psi_{\rm st}= 0$ when applying the equation of motion.

Such a Lagrangian term also exists in the symmetric expression of energy-momentum tensor, which therefore does
not fulfill the constraint in Eq. (\ref{quantum}), either. In fact, the symmetric energy-momentum tensor stands an even
worse situation with respect to such a constraint:
One can check with the familiar electromagnetic field that it does not
guarantee Eq. (\ref{quantum}) even for $i\neq j$.

Our new expression of energy-momentum tensor in Eq. (\ref{Tnew}) does not contain the Lagrangian term.
In the next part of this section, we derive an equivalent and more illuminating expression than Eq. (\ref{Tnew})
to display that such current-correlation property
can be safely guaranteed for a quantum wave in mutual eigen-state of energy and some momentum component.

The  conventional canonical energy-momentum tensor is derived as a N\"other current with the conventional
Lagrangian ${\mathscr  L}_{\rm st}(\varphi,\partial_\mu\varphi)$. From our above discussion, we see that it
almost satisfies the constraint in Eq. (\ref{quantum}), except for the
Lagrangian term which does not in general vanish. Since the Lagrangian of a field can be modified by a surface term
without changing the equation
of motion, this gives us a hint that if we can find a general expression of Lagrangian which always vanishes after
applying the equation of motion,
then the derived N\"other current will automatically satisfy the constraint in Eq. (\ref{quantum}). In what follows, we
show that it is indeed so.

The conventional standard Lagrangians of free scalar, Dirac, and vector fields take the following forms, respectively:
 \begin{subequations}\label{Lst}
\begin{align}
{\mathscr L}_{\rm st}^{\phi}&=-\dfrac{1}{2}\partial^\mu\phi\partial_\mu\phi-\dfrac{1}{2}m^2\phi^2 ,\\
{\mathscr L}_{\rm st}^{\psi}&=\dfrac{1}{2}\overline{\psi}(i\gamma^{\mu}\partial_{\mu}-m)\psi+h.c ,\\
{\mathscr L}_{\rm st}^{A}&=-\dfrac{1}{4}F_{\mu\nu} F^{\mu\nu}-\dfrac{1}{2}m^2A^\mu A_\mu.
\end{align}
\end{subequations}
By noticing that a free-field Lagrangian $ {\mathscr  L}_{\rm st}(\varphi,\partial_\mu\varphi)$ is necessarily quadratic in
the field variable and its derivative, it can be put in a unified form:
 \begin{equation}\label{Lquad}
 {\mathscr  L}_{\rm st}(\varphi,\partial_\mu\varphi)=\dfrac{1}{2}\Big[\varphi\dfrac{\partial{\mathscr  L}_{\rm st}}
{\partial\varphi}
+(\partial_\mu\varphi)\dfrac{\partial{\mathscr  L}_{\rm st}}{\partial(\partial_\mu\varphi)}\Big].
 \end{equation}
By adding a proper surface term, we obtain the desired new expression of Lagrangian
${\mathscr  L}_{\rm new}(\varphi,\partial_\mu\varphi,\partial_\mu\partial_\nu\varphi)$:
\begin{subequations}
\begin{align}
{\mathscr  L}_{\rm new}&={\mathscr L}_{\rm st}-\dfrac{1}{2}\partial_\mu\big[\varphi\dfrac{\partial{\mathscr  L}_{\rm st}}
{\partial(\partial_\mu\varphi)}\big] \\
&=\dfrac{1}{2}\varphi\Big[\dfrac{\partial{\mathscr L}_{\rm st}}{\partial\varphi}
-\partial_\mu\dfrac{\partial{\mathscr L}_{\rm st}}{\partial(\partial_\mu\varphi)}\Big]\label{Lnew},
\end{align}
\end{subequations}
which clearly vanishes by the Euler-Lagrange equation of motion.

The explicit forms of our new Lagrangian for the scalar, Dirac, and vector fields are:
 \begin{subequations}
\begin{align}
\mathscr{L}_{\rm new}^{\phi}&=\frac{1}{2}\phi(\partial_\mu\partial^\mu-m^2)\phi ,\\
\mathscr{L}_{\rm new}^{\psi}&=\frac{1}{2}\overline{\psi}(i\gamma^\mu\partial_\mu-m)\psi+h.c., \\
\mathscr{L}_{\rm new}^{A}&=\frac{1}{2}A_\nu (\partial_\mu F^{\mu\nu}-m^2 A^\nu).
\end{align}
\end{subequations}
For the Dirac field, the ``new'' Lagrangian actually equals the traditional expression, which is already zero
by the equation of motion.

Notice that the new Lagrangian ${\mathscr  L}_{\rm new}$ contains a second derivative,
thus the derivation of N\"other current is a little bit (but not much) more involved \cite{Lomp14}. The result is:
 \begin{equation}\label{TnewH}
T^{\mu\nu}_{\rm new}=-i\big[\frac{\partial{\mathscr  L}_{\rm new}}{\partial(\partial_{\mu}\phi_{a})}+\frac{\partial{\mathscr
L}_{\rm new}}{\partial(\partial_{\mu}\partial_{\sigma}\phi_{a})}\partial_{\sigma}-\partial_{\sigma}\frac{\partial{\mathscr
L}_{\rm new}}{\partial(\partial_{\sigma}\partial_{\mu}\phi_{a})}\big] {\textbf{P}}^{\nu} \phi_{a},
\end{equation}
where $\textbf{P}^{\nu}=-i\partial^{\nu}$ is the quantum-mechanical four-momentum operator. We call this a ``hyper-canonical'' form, as it is a single expression with the single operator inserted for the
desired observable. Such a hyper-canonical form clearly guarantees the current-correlation property as we
elaborated above for a quantum wave in mutual eigen-state of two or more components of $\textbf{P}^\nu$.

By a slight algebra, Eq. (\ref{TnewH}) can be converted into the more convenient expression with the conventional Lagrangian ${\mathscr L}_{\rm st}$ containing only the first derivative:
\begin{equation}
T^{\mu\nu}_{\rm new}=-\frac{\partial\mathscr{L}_{\rm st}}{\partial(\partial_\mu\varphi)}\overleftrightarrow{\partial}^\nu\varphi
=-\frac{1}{2} \big( \frac{\partial\mathscr{L}_{\rm st}}{\partial(\partial_\mu\varphi)}{\partial}^\nu\varphi
-\partial^\nu\frac{\partial\mathscr{L}_{\rm st}}{\partial(\partial_\mu\varphi)}\varphi \big).
\end{equation}

If one considers further the measurement of angular momentum, a proper expression satisfying the correct correlation
of currents can also be derived:
\begin{eqnarray}\label{Mnew}
M^{\mu\nu\rho}_{\rm new}&=&\big(x^\mu T_{\rm new}^{\rho\nu}-x^\nu T_{\rm new}^{\rho\mu}\big)
+\big(-i\frac{\partial\mathscr{L}_{\rm new}}{\partial(\partial_\rho\varphi_a)} \Sigma ^{\mu\nu}_{ab}\varphi_b
-\frac{1}{2}\eta^{\rho\nu}\frac{\partial\mathscr{L}_{\rm new}}{\partial(\partial_\mu\varphi_a)}\varphi_a
+\frac{1}{2}\eta^{\rho\mu}\frac{\partial\mathscr{L}_{\rm new}}{\partial(\partial_\nu\varphi_a)}\varphi_a \big)
\nonumber \\
&\equiv& L^{\mu\nu\rho}_{\rm new}+s^{\mu\nu\rho}_{\rm new}.
\end{eqnarray}

In the next section, we wish to explore what theory we get by setting the new energy-momentum tensor as the gravitational source.

\section{The gravitational coupling from the new energy-momentum tensor}
The simplest way to switch on the gravitational interaction is by minimal coupling, namely, replacing the Lorentz-covariant
quantities in the flat space-time action of matter fields with their counterparts which are covariant under arbitrary coordinate transformation. Let us first make this attempt in the Riemann space-time, starting with the Lagrangian in Eq. (\ref{Lnew}) which can give the desired N\"other currents:
\begin{equation}\label{Rnew}
\mathscr{L}_{\rm new}(\varphi,\partial\varphi,\partial^2\varphi)
\rightarrow \sqrt{g}\widetilde{L}_{\rm new} (\varphi,\widetilde{\nabla}\varphi,\widetilde{\nabla}\widetilde{\nabla} \varphi)
=\sqrt{g}\widetilde{L}_{\rm st} (\varphi,\widetilde{\nabla}\varphi)-\dfrac{1}{2}\sqrt{g}\widetilde{\nabla}_\mu\big[\varphi\dfrac{\partial  \widetilde{L}_{\rm st}}{\partial(\widetilde{\nabla}_\mu\varphi)}\big],
\end{equation}
where $g$ is the determinant of the metric tensor, and $\widetilde{X}$ denotes a quantity $X$ in the Riemann space-time.
For example,
$\widetilde{\nabla}_{\mu}$ signifies the covariant derivative built with Christoffel connection.
However, nothing new will be gained in the Riemann space-time, in which the covariant Gauss' theorem tells us that the surface term in Eq. (\ref{Rnew}) is irrelevant:
\begin{equation}\label{Inew}
\dfrac{1}{2}\sqrt{g}\widetilde{\nabla}_\mu\big[\varphi\dfrac{\partial\widetilde{L}_{\rm st}}{\partial(\widetilde{\nabla}_\mu\varphi)}\big]
=\partial_\mu\big[\dfrac{1}{2}\sqrt{g}\varphi\dfrac{\partial\widetilde{L}_{\rm st}}{\partial(\widetilde{\nabla}_\mu\varphi)}\big].
\end{equation}

So let us try instead the Riemann-Cartan space-time with non-zero torsion. To facilitate the inclusion of spinor field, we use the tetrad formalism \cite{Hehl76,Hehl17}, in which the gravitational theory is described by an action principle of the form
\begin{equation}
I({e^a}_\mu,{\omega^a}_{b\mu},\varphi)={\int}d^{4}xe[\frac{1}{16\pi{G}}
R({e^a}_\mu,{\omega^a}_{b\mu})+L({e^a}_\mu,{\omega^a}_{b\mu},\varphi)],
\end{equation}
where R is the curvature scalar with torsion, e is the determinant of the tetrad ${e^a}_\mu$, and
$L$ is the lagrangian for matter. In this paper,
Latin and Greek letters denote Lorentz and coordinate indices, respectively. The tetrad quantities are defined by
$g_{\mu\nu}={e^a}_\mu {e^b}_\nu\eta_{ab}$, $\eta_{ab}={e_a}^\mu {e_b}^\nu g_{\mu\nu}$, and ${\omega^a}_{b\mu}$ is the spin connection as appear in the covariant derivative, e.g., of a Lorentz-vector
$A^a={e^a}_\mu A^\mu$:
$\nabla_\mu A^a=\partial_\mu A^a+{\omega^a}_{b\mu}A^b$.

When varying the total action of matter and geometry $I=I_g+I_m $ with respect to tetrad and spin connection independently,
one can obtain equations of motion of gravitation:
\begin{equation}
{G^\mu}_a=8\pi{G}{\mathbf{T}^\mu}_a,~
{\mathcal S _a}^{b\mu}=-4\pi{G}{\mathbf{s}_a}^{b\mu}.
\end{equation}
The Einstein tensor $G^{\mu\nu}=e^{a\nu}{G^\mu}_a$ is generally asymmetric.
${\mathcal S}^{\mu\nu\rho}(=e^{a\mu}{e_b}^\nu {{\mathcal S}_a}^{b\rho}=S^{\mu\nu\rho}+g^{\mu\rho}S^\nu-g^{\nu\rho}S^\mu)$
is called the modified torsion tensor, and the torsion tensor is
$ {S_{\mu\nu}}^\rho=\Gamma_{[\mu\nu]}^\rho={\omega^\rho}_{[\mu\nu]}+{e_a}^\rho\partial_{[\nu}{e^a}_{\mu]}$.
Its trace ${S_{\mu\nu}}^\nu \equiv S_\mu$ is defined as the torsion vector. The energy-momentum tensor and
spin tensor can be conveniently evaluated as
\begin{equation}\label{tetradTS}
\mathbf{T}^{\mu\nu} \equiv
e^{a\nu} \cdot \big( {\mathbf{T}^\mu}_a \equiv  \frac{1}{e}\frac{\delta I_m}{\delta{e^a}_\mu} \big),
~
\mathbf{s}^{\mu\nu\rho} \equiv
e^{a\mu}{e_b}^\nu \cdot \big( {\mathbf{s}_a}^{b\rho} \equiv -\frac{2}{e}\frac{\delta I_m}{\delta{\omega^a}_{b\mu}}\big).
\end{equation}

In the Einstein-Cartan theory, the matter Lagrangian $L_{\rm st}$ is constructed via minimal coupling of the standard Lagrangian:
\begin{equation}\label{Nnew}
\mathscr{L}_{\rm st}(\varphi,\partial\varphi)
\rightarrow eL_{\rm st} (e,\omega,\varphi,\nabla\varphi),
\end{equation}
and presets the conventional cannoical energy-momentum and spin tensors as the source of gravity. In our model, it is constructed via minimal coupling of the Lagrangian in Eq. (\ref{Lnew}):
\begin{equation}\label{Nnew}
\mathscr{L}_{\rm new}(\varphi,\partial\varphi,\partial^2\varphi)
\rightarrow eL_{\rm new} (e,\omega,\varphi,\nabla\varphi,\nabla\nabla \varphi)
=eL_{\rm st} (e,\omega,\varphi,{\nabla}\varphi)-\dfrac{1}{2}e{\nabla}_\mu\big[\varphi\dfrac{\partial{ L}_{\rm st}}{\partial({\nabla}_\mu\varphi)}\big].
\end{equation}
With $\nabla_{\mu} A^{\mu}=\widetilde{\nabla}_{\mu} A^{\mu}+2S_{\mu} A^{\mu}$ \cite{Hehl76}, the last term in Eq. (\ref{Nnew}) can be expressed as:
\begin{equation}
\dfrac{1}{2}e{\nabla}_\mu\big[\varphi\dfrac{\partial L_{\rm st}}{\partial({\nabla}_\mu\varphi)}\big]
= \dfrac{1}{2}e\widetilde{\nabla}_\mu\big[\varphi\dfrac{\partial L_{\rm st}}{\partial({\nabla}_\mu\varphi)}\big]
+eS_\mu\varphi\dfrac{\partial L_{\rm st}}{\partial({\nabla}_\mu\varphi)}
=\partial_\mu\big[\dfrac{1}{2}e\varphi\dfrac{\partial L_{\rm st}}{\partial(\widetilde{\nabla}_\mu\varphi)}\big]
+eS_\mu\varphi\dfrac{\partial L_{\rm st}}{\partial({\nabla}_\mu\varphi)}.
\end{equation}
It's not hard to see that the equivalence between $\mathscr{L}$ and $\mathscr{L}_{\rm new}$ in flat or even Riemann space-time
is lost in the  Riemann-Cartan space-time. The difference between the matter action $I^{\rm new}_m(e, \omega, \varphi)$ in our model and that in the Einstein-Cartan theory is
\begin{eqnarray}\label{compare}
I^{\rm new}_m(e, \omega, \varphi)=
{\int}d^{4}xeL_{\rm new}(\varphi,\nabla\varphi,\nabla\nabla \varphi)
=I_m^{\rm EC}-{\int}d^{4}xe \varphi\frac{\partial{L}}{\partial(\nabla_{\mu}\varphi)}S_{\mu}.
\end{eqnarray}

We will show in the next section with the specific scalar, spinor, and vector fields that in our model the
energy-momentum tensor is exactly the covariant extension of $T_{\rm new}^{\mu\nu}$ in Eq. (\ref{Tnew}).
Thus, guided by the new energy-momentum tensor, we can indeed arrive at a sensible model of gravitational interaction,
derived by minimal coupling of a specific and valid Lagrangian in flat space-time.

It should be remarked that our model can also be viewed as one with minimal coupling.
The point is that the equivalence between $\mathscr{L}_{\rm st}$ and $\mathscr{L}_{\rm new}$ in flat or even Riemann space-time
is lost in the  Riemann-Cartan space-time. If one regards ${L}_{\rm new}$ as the fundamental Lagrangian, our model is a minimal-coupling one.
On the other hand, if priority is assigned to ${L}_{\rm new}$ which contains only the first derivative, then
as the last expression of Eq. (\ref{Inew}) indicates, our model can be viewed as an extension of Einstein-Cartan
theory by including a non-minimal coupling between matter and torsion.
Such ambiguity of minimal coupling in the Einstein-Cartan space-time had already been discussed before
\cite{Kazm08, Kazm09, Kazm09'}. From our analysis in this paper,
even the original Einstein-Cartan action can be regarded as one of non-minimal coupling,
were  our ${L}_{\rm new}$ given a priority.

\section{Explicit construction for scalar, spinor, and vector fields}
In this section, we present explicitly the new gravitational coupling for the scalar, spinor, and vector fields,
respectively, and verify that the gravitational source is indeed the covariant extension of the energy-momentum tensor and spin tensor currents as
required to properly describe the fluxes of conserved quantities in quantum measurement.

\subsection{Scalar field}\label{scalar}
The proper flat space-time Lagrangian for the free scalar field $\phi$ in the form of Eq. (\ref{Lnew}) is:
 \begin{equation}
{\mathscr L}_{\phi }^{\rm new}=\phi(\partial_\mu\partial^\mu-m^2)\phi.
\end{equation}
In the spirit of Eq. (\ref{compare}), the corresponding action in Riemannian-Cartan space-time is
\begin{subequations}
\begin{align}
I_\phi^{\rm new} (e,\omega,\phi)&={\int}d^{4}xe\frac{1}{2}[\phi(\nabla_a\nabla^a-m^2)\phi]
\equiv {\int}d^{4}xe {L}_{\phi }^{\rm new} (e,\omega,\phi)\\
&={\int}d^{4}xe[-\frac{1}{2}\nabla_a\phi\nabla^a\phi-\frac{1}{2}m^2\phi^2
+\phi(\nabla_a\phi) S^a]
\equiv {\int}d^{4}xe {L}_{\phi }^{\rm 'new} (e,\omega,\phi).
\end{align}
\end{subequations}
As we noted above, the first expression is of the form of minimal coupling, while the second expression
displays a non-minimal coupling, which indicates clearly that the scalar field interacts with torsion. This
action gives the equation of motion for $\phi$:
\begin{equation}\label{phi-our}
\frac{1}{2}\nabla_a\nabla^a\phi+\frac{1}{2}\stackrel{\ast~}{\nabla_a}\stackrel{\ast~}{\nabla^a}\phi-m^2\phi=0,
\end{equation}
where $\stackrel{\ast~}{\nabla_a}=\nabla_a-2S_a, \stackrel{\ast~}{\nabla^a}=\nabla^a-2S^a$ is called the modified divergence. It compares to
\begin{equation}\label{phi-EC}
\stackrel{\ast~}{\nabla_a}{\nabla^a}\phi-m^2\phi=0
\end{equation}
in the Einstein-Cartan theory with the action
\begin{equation}
I_\phi (e,\omega,\phi)={\int}d^{4}xe[-\frac{1}{2}\nabla_a\phi\nabla^a\phi-\frac{1}{2}m^2\phi^2] .
\end{equation}

By computing Eq. (\ref{tetradTS}) explicitly and applying the covariant equation of motion, we get
the tetrad energy-momentum tensor and tetrad spin tensor of the scalar field:
\begin{equation}\label{Tphi}
\mathbf{T}^{\mu\nu}=\frac{1}{2}\stackrel{\ast~}{\nabla^\mu}\phi\nabla^\nu\phi-\frac{1}{2}\phi\nabla^\nu\nabla^\mu\phi
+g^{\mu\nu}L_\phi^{\rm new},
~
\mathbf{s}^{\mu\nu\rho}=\frac{1}{2}(g^{\nu\rho}\phi\nabla^\mu\phi-g^{\mu\rho}\phi\nabla^\nu\phi).
\end{equation}
In the flat space-time limit, they reduce to the forms in Eqs. (\ref{Tnew}) and (\ref{Mnew}):
\begin{equation}\label{phi-flat}
\mathbf{T}_{(0)}^{\mu\nu}=T^{\mu\nu}_{\rm eff}=\partial^\mu\phi\overleftrightarrow{\partial}^\nu\phi,
~
\mathbf{s}_{(0)}^{\mu\nu\rho}={s}^{\mu\nu\rho}_{\rm eff}=\eta^{[\nu|\rho}\phi\partial^{|\mu]}\phi.
\end{equation}

The last term in Eq.(\ref{Tphi}) and Eq.(\ref{phi-flat}) might be regarded as an extra "spin" current, or more exactly a "pseudo-spin" current, as this current does not contribute to the component $M^{0ij}_{new}$ and thus does not contribute to the integrated spin "charge". We call it "spin" just because it is anyway not an orbital type, it can join the traditional spin current to couple to the space-time torsion in a modified gravitational model.

\subsection{Spinor field}
For the Dirac spinor field, our model actually coincides with the Einstein-Cartan theory. The reason is that the usual
expression of Dirac Lagrangian in flat space-time:
 \begin{equation}
{\mathscr L} _\psi =\frac{i}{2}\overline{\psi}\gamma^\mu\partial_\mu\psi
-\frac{1}{2}m\overline{\psi}\psi+h.c.,
\end{equation}
is itself zero when applying the Dirac equation, thus is already of the form in Eq.(\ref{Lnew}). So, when going to
Riemannian-Cartan space-time, our approach gives the same action of the Einstein-Cartan type:
\begin{equation}
I_\psi(e,\omega,\psi,\overline{\psi})={\int}d^{4}x e
\big( \frac{i}{2}\overline{\psi}\gamma^a\nabla_a\psi-\frac{1}{2}m\overline{\psi}\psi+h.c.\big )
\equiv {\int}d^{4}x e L_\psi,
\end{equation}
where $\nabla_a\psi={e_a}^\mu(\partial_\mu+\frac{i}{4}\sigma_{ab}{\omega^{ab}}_\mu)\psi.$
The equations of motion are
\begin{equation}\label{psi-EC}
\frac{i}{2}\gamma^a\nabla_a\psi+\frac{i}{2}\gamma^a\stackrel{\ast~~}{\nabla_a}\psi-m\psi=0,
~~
\frac{i}{2}\nabla_a\overline{\psi}\gamma^a+\frac{i}{2}\stackrel{\ast~~}{\nabla_a}\overline{\psi}\gamma^a+m\overline{\psi}=0
\end{equation}
which resemble more of Eq.  (\ref{phi-our}) in our model, rather than Eq.  (\ref{phi-EC}) in the Einstein-Cartan theory.

The tetrad energy-momentum tensor and tetrad spin tensor of the Dirac field are trivial covariant extension
of the familiar canonical expressions:
\begin{equation}
\mathbf{T}^{\mu\nu}=-\frac{i}{2}\overline{\psi}\gamma^\mu\nabla^\nu\psi+h.c. +g^{\mu\nu} L_\psi,
~
\mathbf{s}^{\mu\nu\rho}=
\frac{1}{4}\overline{\psi}(\gamma^\rho\sigma^{\mu\nu}+\sigma^{\mu\nu}\gamma^\rho)\psi.
\end{equation}

\subsection{Vector field}
For a massive vector field, the proper flat space-time Lagrangian in the form of Eq. (\ref{Lnew}) is:
 \begin{equation}
{\mathscr L}_A^{\rm new}=\frac 12A_\nu(\partial_\mu F^{\mu\nu}-m^2A^\nu).
\end{equation}
Following again Eq. (\ref{compare}), we get the corresponding action in Riemann-Cartan space-time:
\begin{subequations}
\begin{align}
I_A^{\rm new}(e,\omega,A)&=\frac{1}{2}{\int}d^{4}x e[A_b(\nabla_a F^{ab}-m^2A^b)]
\equiv {\int}d^{4}x e L_A^{\rm new}(e,\omega,A)
\\
&={\int}d^{4}x e(-\frac{1}{4}F^{ab}F_{ab}-\frac{1}{2}{m^2}A^2+A^{b}F_{ab}S^{a})
\equiv {\int}d^{4}x e L_A^{\rm 'new}(e,\omega,A) .
\end{align}
\end{subequations}
Viewed either as a minimal-coupling one or non-minimal-coupling one, this action gives a spin-torsion
interaction of both the traditional type as in Einstein-Cartan theory, and a new type related to the
extra spin current in Eq. (\ref{Mnew}). It gives the equation of motion:
\begin{equation}\label{A-our}
\frac{1}{2}\nabla_a F^{ab}+\frac{1}{2}\stackrel{\ast~}{\nabla_a}\stackrel{\ast~~}{F^{ab}}-m^2A^b=0,
\end{equation}
compared to
\begin{equation}\label{A-EC}
I_A(e,\omega,A)={\int}d^{4}x e(-\frac{1}{4}F^{ab}F_{ab}-\frac{1}{2}{m^2}A^2),~~
\stackrel{\ast~}{\nabla_a}{F^{ab}}-m^2A^b=0
\end{equation}
in the Einstein-Cartan theory.  Again, it is Eq. (\ref{A-our}) of our model rather than Eq. (\ref{A-EC}) of the original
Einstein-Cartan theory that resembles more of Eq. (\ref{psi-EC}). Together with the similarity between Eqs. (\ref{phi-our})
and (\ref{psi-EC}), this might be regarded as a hint that our chosen
Lagrangian in Eq. (\ref{Lnew}) is better for minimal coupling in the Einstein-Cartan space-time.

Inserting our new action into Eq. (\ref{tetradTS}) and applying the covariant equation of motion,
we get the tetrad energy-momentum tensor and tetrad spin tensor of the massive vector field:
\begin{equation}\label{TA}
\mathbf{T}^{\mu\nu}=\frac{1}{2}\stackrel{\ast~~}{F^{\mu\rho}}\nabla^\nu A_\rho-\frac{1}{2}\nabla^\nu F^{\mu\rho}A_\rho
+g^{\mu\nu}L_A^{\rm new},
~
\mathbf{s}^{\mu\nu\rho}
=A^{[\mu} F^{\nu]\rho}+A^{[\mu}\stackrel{\ast~~~}{F^{\nu]\rho}}+g^{[\nu|\rho} A_\lambda F^{|\mu]\lambda},
\end{equation}
where $\stackrel{\ast~~}{F^{\mu\nu}}\equiv \stackrel{\ast~}{\nabla^\mu} A^\nu-\stackrel{\ast~}{\nabla^\nu} A^\mu$.

Again, in the flat space-time limit, they reduce to the forms in Eqs. (\ref{Tnew}) and (\ref{Mnew}):
\begin{equation}
\mathbf{T}_{(0)}^{\mu\nu}=T^{\mu\nu}_{\rm eff}=F^{\mu\rho}\overleftrightarrow{\partial}^{\nu}A_\rho,
~
\mathbf{s}_{(0)}^{\mu\nu\rho}={s}^{\mu\nu\rho}_{\rm eff}=2A^{[\mu|}F^{|\nu]\rho}+\eta^{\rho[\nu|}F^{|\mu]\alpha}A_\alpha.
\end{equation}

\section{Alternative theory for the scalar field}

It is notable that although the scalar field in our model acquires a spin current, its new
energy-momentum tensor remains symmetric. This suggests that we may conjecture a model
within the Riemann space-time to assign our new energy-momentum tensor as the gravitational source.
This goal, however, can only be possibly achieved with a substantially non-minimal coupling, which cannot be
converted into a total divergence and thus can survive in the Riemann space-time.
We find that the following action makes such a model:
\begin{equation}
I(g,\phi)=I_g(g)+I_{\phi}(g,\phi)={\int}d^{4}x\sqrt{g} \frac{\widetilde{R}}{16\pi G}
+{\int}d^{4}x\sqrt{g}(-\frac{1}{2}\widetilde{\nabla}_\mu\phi\widetilde{\nabla}^\mu\phi
-\frac{1}{2}m^2\phi^2+\frac{1}{8}\widetilde{R}\phi^2).
\end{equation}
Here $\widetilde{X}$ denotes a quantity $X$ in the Riemann space-time.
For example, $\widetilde{\Gamma}_{\mu\nu}^\rho$ signifies the Christoffel connection.
The action contains a non-minimal $\widetilde{R}\phi^2$ interaction, with the coupling coefficient fixed to be 1/8.
It gives an energy-momentum tensor of the scalar field
 \begin{equation}
 \mathbf{T}^{\mu\nu}=\frac{2}{\sqrt{g}}\frac{\delta{I_{\phi}(g,\phi)}}{\delta{g_{\mu\nu}}}
 =\frac{1}{2}(\widetilde{\nabla}^\mu\phi\widetilde{\nabla}^\nu\phi-\phi\widetilde{\nabla}^\mu\widetilde{\nabla}^\nu\phi)
-\frac{1}{4}\widetilde{R}^{\mu\nu}\phi^2.
\end{equation}
In the limit of flat space-time, this agrees with Eq. (\ref{phi-flat}) of Section \ref{scalar}.

Note that the scalar field here is not necessarily a fundamental field as employed in cosmological models.
It can as well describe a spin-less composite particle.

\section{Discussion}
As the Einstein-Cartan theory does, our model leads to spin contact interaction, but of more extensive structures.
Certainly,  the test of such contact interaction has to await extremely precise measurements \cite{Shap98,Obuk14,Duan16},
but it does not mean that our discussion is purely academic.
In fact, the most valuable light which our study might shed on the test of gravitational effect,
independent of the possible merit of our gravitational-coupling model itself,
is that if $T_g^{\mu\nu}$ differs from $T_{\rm new}^{\mu\nu}$,  then some peculiar effect may occur.
For example, in Einstein's general relativity, it is $T_g^{\mu\nu}$ that couples to
gravity, while during a quantum measurement
the effective fluxes of energy and momentum of a quantum wave is dictated by
$T_{\rm new}^{\mu\nu}$ in Eq. (\ref{Tnew}), which is indeed different from $T_g^{\mu\nu}$.
This may lead to some kind of non-local violation of universality of free fall for quantum waves,
and one may conjecture a possible gravitational discrimination of freely falling atomic waves of different species.

In this paper, we have worked with massive vector field to avoid the discussion of gauge invariance, which is highly tricky
and controversial \cite{Lead14}.  In the Riemann-Cartan space-time, the minimal coupling between gauge field
and torsion is gauge-dependent and hence often abandoned \cite{Hojm78}.
Nethertheless, the recent technique to construct gauge-invariant gluon spin
\cite{Lead14, Chen08, Chen09}
may be adopted to build a gauge-invariant minimal coupling of photon or gluon to torsion.
Thus, exploring the interaction between torsion and gauge particles is of vital importance and interest for
the fundamental aspects of not only gravity, but also gauge theory.

We thank Wei Xu and De-Tian Yang for helpful discussion.
This work is supported by the China NSF via Grants No. 11535005 and No. 11275077.

\end{document}